\documentclass[12pt,preprint]{aastex} 

\begin{document}

\shortauthors{Hunter et al.}

\shorttitle{Holography of the GBT}

\title{Holographic Measurement and Improvement of the Green Bank Telescope
   Surface}

\author{Todd~R.~Hunter,\altaffilmark{1}
  Frederic~R.~Schwab,\altaffilmark{1} 
  Steven~D.~White,\altaffilmark{2} 
  John~M.~Ford,\altaffilmark{2}  
  Frank~D.~Ghigo,\altaffilmark{2} 
  Ronald~J.~Maddalena,\altaffilmark{2}  
  Brian~S.~Mason,\altaffilmark{1}  
  Jack~D.~Nelson,\altaffilmark{2}
  Richard~M.~Prestage,\altaffilmark{2}
  Jason~Ray,\altaffilmark{2}  
  Paul~Ries,\altaffilmark{2,4}  
  Robert~Simon,\altaffilmark{2} 
  Sivasankaran~Srikanth,\altaffilmark{3}
  Peter~Whiteis\altaffilmark{2}}
\email{thunter@nrao.edu}

\altaffiltext{1}{National Radio Astronomy Observatory, 520 Edgemont Rd., 
   Charlottesville, VA, 22903} 
\altaffiltext{2}{National Radio Astronomy Observatory, P.O. Box 2, 
   Green Bank, WV, 24944} 
\altaffiltext{3}{National Radio Astronomy Observatory, 
   1180 Boxwood Estate Rd., Charlottesville, VA, 22903}
\altaffiltext{4}{University of Virginia, Astronomy Dept., P.O. Box
   3818, Charlottesville, VA, 22903}

\begin{abstract}
We describe the successful design, implementation, and operation of a
12-GHz holography system installed on the Robert C. Byrd Green Bank
Telescope (GBT)\@.  We have used a geostationary satellite beacon to
construct high-resolution holographic images of the telescope mirror
surface irregularities.  These images have allowed us to infer and apply
improved position offsets for the 2209 actuators which control the active
surface of the primary mirror, thereby achieving a dramatic reduction in
the total surface error (from 390$\,\mu$m to ${\sim}240\,\mu$m, rms).  We
have also performed manual adjustments of the corner offsets for a few
panels.  The expected improvement in the radiometric aperture efficiency
has been rigorously modeled and confirmed at 43~GHz and 90~GHz. The
improvement in the telescope beam pattern has also been measured at
11.7~GHz with greater than 60~dB of dynamic range.  Symmetric features in
the beam pattern have emerged which are consistent with a repetitive
pattern in the aperture due to systematic panel distortions.  By computing
average images for each tier of panels from the holography images, we
confirm that the magnitude and direction of the panel distortions, in
response to the combination of gravity and thermal gradients, are in
general agreement with finite-element model predictions.  The holography
system is now fully integrated into the GBT control system, and by
enabling the telescope staff to monitor the health of the individual
actuators, it continues to be an essential tool to support high-frequency
observations.
\end{abstract}


\section{Introduction}

The Robert C. Byrd Green Bank Telescope (GBT) of the National Radio
Astronomy Observatory (NRAO) in Green Bank, WV, is a 100-meter
fully-steerable single dish radio telescope operating at frequencies from
290~MHz to 100~GHz.  The GBT uses a dual-offset Gregorian design providing
a fully unblocked aperture.  With a total moving mass of 7700 metric tons
\citep{Parker05}, its design and operational status as of 2008 were
summarized by \citet{Prestage09}.  The research program to further improve
the high-frequency performance of the telescope is known as the Precision
Telescope Control System (PTCS)\@.  This program has fielded several
successful innovations in recent years.  In addition to an advanced
pointing model including real-time thermal correction terms
\citep{Constantikes07,Constantikes04}, an optical quadrant detector is now
in use by observers working at the highest frequency bands, in order to
measure and correct for dynamic pointing errors due to wind-induced motion
of the feedarm \citep{Ries11}. Also, efforts are well underway to
implement an advanced digital servo control system to achieve subarcsecond
tracking performance.  The other major goal of the PTCS work has been to
improve the surface accuracy of the GBT primary mirror.

In order to achieve and maintain an accurate paraboloidal shape, the GBT
primary was built with an active surface comprising 2004 panels and 2209
actuators.  The geometry of the surface is described by \citet{Schwab90},
and the hardware and control system is described by \citet{Lacasse98}.
The actuator home positions were set originally (in Spring 2000) via
optical photogrammetry of targeted points on the surface, with the
telescope parked at a single elevation.  In 2002, an elevation-dependent
finite element model (FEM) was incorporated into the active surface
control system in order to achieve a more uniform gain as a function of
elevation.  The residual errors in the FEM correction were later measured
via the out-of-focus (OOF) holography technique at 43~GHz
\citep{Nikolic07a}.  The residuals were modeled using a set of
elevation-dependent Zernike polynomial expansions, and the subsequent
application of the associated Zernike-based corrections resulted in an
essentially flat gain curve at 43~GHz \citep{Nikolic07b}.  At this point,
the measured aperture efficiency corresponded to a surface error of about
390\,$\mu$m rms.  The GBT monitor and control (M\&C) system now implements
these corrections automatically at the beginning of each scan during all
high-frequency observations.  Since the measured panel manufacturing error
was only 75\,$\mu$m rms, and a model of the thermal plus gravitational
error of the individual panels was computed to be $<125\,\mu$m rms
\citep{RSI92}, the major contributor to the total surface error was
believed to be errors in the actuator zero positions derived from
photogrammetry.

In late Fall 2008, we installed a Ku-band (12-GHz) holography system on
the telescope in order to measure the residual errors in the actuator
zero positions.  By obtaining large maps of the telescope's beam pattern,
the technique of ``with-phase'' holography allows one to measure the
complex electric field distribution across the aperture and subsequently
use the phase distribution to construct a high-resolution image of the
surface irregularities \citep[e.g.,][]{Bennett76,Scott77}.  These images
can then be used to derive corrections for adjusting the panels
mechanically to reduce the alignment errors.  In this paper, we describe
the instrument and present the results of the observations and the
surface correction strategy that we followed during 2009--2010.  We also
review the models of systematic panel-scale errors and compare them to the
observed tier-averaged profiles from the holography maps. Finally, we
discuss the effect of thermal environment on the ultimate surface
performance of the GBT.

\section{Instrumentation}

As with similar systems deployed on other telescopes
\citep[e.g.][]{Baars07,Balasubramanyam09,Grahl86}, the holography system
for the GBT is composed of two receivers.  Originally designed in the
early 1990s, each holography receiver consists of a low-noise block
downconverter (LNB) and an external phase-locked local oscillator (LO)
that provides the initial downconversion of signals from the Ku satellite
downlink band (11.7--12.2 GHz) to an intermediate frequency (IF) range of
0.95 to 1.45 GHz.  This IF signal can be sent to any of the GBT backends
in the Jansky Laboratory via the normal fiber-optic transmission system.
We modified both LNBs to accept the LO signal (10.7~GHz) from an external
dielectric resonator oscillator (DRO).  The DROs are digitally
phase-locked and deliver superior phase stability over long time periods.

The primary holography receiver illuminates the subreflector from a
standard receiver slot in the Gregorian focus turret, allowing it to
remain on the telescope for a large fraction of the year.  The feedhorn
illumination of the subreflector is weakly tapered ($-4$~dB), compared to
the typical edge taper value of $-13$~dB used with the standard astronomy
receivers to achieve low sidelobe levels and minimal spillover loss.  This
shallower taper maintains sensitivity to surface features near the edge of
the primary dish.  The reference receiver is coupled to an upward-looking
30-cm diameter feedhorn located at the tip of the vertical feed arm above
the subreflector.  The 3-dB beamwidth of this receiver as measured on the
sky is 6.4\arcdeg\ in elevation and 6.5\arcdeg\ in azimuth, and its
direction of peak response is aligned parallel to within 0.9\arcdeg\ of
the main telescope axis. For this receiver, the LO and IF signals are sent
through phase-stable cable to the receiver cabin.

The holography backend is a digital complex correlator that resides in the
receiver cabin.  The two-channel IF processor consists of a bandpass
filter at \hbox{50.0--50.1~MHz}, a programmable attenuator, a single
sideband (SSB) downconverter, and an anti-aliasing filter with a 3-dB
bandwidth of 8.8~kHz.  The signal channel for the receiver illuminating
the main dish also includes a 90\arcdeg\ phase shifter (Hilbert transform
network) accurate to $\pm 1$\arcdeg\ across the 10-kHz bandwidth.  The
three IF outputs are digitized at 24~kHz by 16-bit analog-to-digital
converters (ADCs) from which the correlator calculates six products: the
autocorrelation of the signal from the main dish $T\star T$, the
autocorrelation of the signal from the reference antenna $R\star R$, the
autocorrelation of the phase shifted signal from the main dish $Q\star Q$,
and the three corresponding cross-correlations $(R\star T$, $R\star Q$,
and $T\star Q)$.  The correlated signal amplitude and phase are derived
from $R\star Q$ and $R\star T$; the other four products are useful for
realtime observing diagnostic tests.

For our application, we need to measure the complex voltage reception
pattern at sub-meter resolution across the 100-m aperture with a noise
level in the phase corresponding to $<100\,\mu$m (half path).  To achieve
this performance at 12~GHz requires imaging a point source over a
2\arcdeg~field with a signal-to-noise ratio of ${\sim}35$~dB in voltage
(${\sim}70$~dB in power) \citep{Scott77,Rochblatt91}.  Thus, the receiver
requires an abnormally large dynamic range, which is probably the highest
that has ever been needed for a radio telescope holography experiment.
Great care was taken in the amplifier chain to preserve a large dynamic
range, and after some necessary modifications it was measured to be
${\sim}72$~dB\@.  During Summer 2008, end-to-end laboratory tests using
injected signals of the same strength expected from a satellite showed a
system phase stability of 0.4\arcdeg\ rms in 36-millisecond integrations.
In terms of surface error, 0.4\arcdeg\ of phase equates to
${\sim}60\,\mu$m and this indicates that the receiver performance should
be adequate for the task.

\section{Observational Technique}

\subsection{Satellite targets}

Beginning in August 2008, we used the GBT cryogenic Ku-band receiver to
perform a spectral survey of a few dozen geosynchronous satellites visible
from Green Bank.  The GBT is capable of following two-line element (TLE)
sets that describe the Keplerian orbital elements of a satellite.  We use
the TLEs published continuously online by Dr.\
T.~S.~Kelso\footnote{http://celestrak.com/NORAD/elements}.  In our survey,
we identified a number of unmodulated, strong, and stable continuous-wave
beacons at or near 11.700 GHz that are suitable for holographic mapping.
Mostly these are Intelsat satellites in the Galaxy series.  Due to its
proximity to the longitude of Green Bank, the primary target became
Galaxy~28 at 89\arcdeg\ west longitude.  Launched in June 2005, it appears
at an elevation of 44\arcdeg\ from Green Bank.  The other target we have
occasionally used is Galaxy~27 at 129\arcdeg\ west longitude and appearing
at an elevation of 27\arcdeg\ from Green Bank.

After installing the holography system in December 2008, we measured the
typical phase stability of the system to be between 1--2\arcdeg\ rms in
36-millisecond integrations in good weather conditions.  While this value
is a factor of a few larger than the values measured in the lab, it is
consistent with the best quartile of the historical data from a 100-m
baseline atmospheric phase monitor \citep{Radford96} which operated at the
GBT site from 2000 to 2004 \citep{Maciolek00}. The frequency of the beacon
was found to be stable throughout the day and night to much less than the
anti-aliasing filter bandwidth; thus after centering the signal once for a
particular satellite, no further adjustment was necessary.

\subsection{Pointing, focusing, and mapping}

We began the holography campaign in earnest in January 2009.  Each session
begins by taking pointing cross-scans of the main dish across the
satellite target using the full continuum signal routed to the Digital
Continuum Receiver (DCR) backend. The pointing corrections are determined
automatically by ASTRID \citep{Oneil06}.  Often the pointing requires a
few cycles as the TLEs are generally only accurate to about 0.5--1\arcmin,
whereas the GBT beamsize is 1\arcmin.  A focus scan to find the proper
focus offset is also measured on the satellite.  Maps are acquired using
on-the-fly raster scanning \citep{Mangum2007} over a
$2^\circ\times2^\circ$ region with $\sim$1400 points in the scan direction
and 201 points in the perpendicular direction, requiring approximately 3.5
hours.  Every 15~minutes or so, we return to the position of the satellite
and integrate for $\sim$1 minute to obtain what we term a reference scan.
We examine the reference scans immediately in order to detect the
magnitude of any pointing drift and to monitor the atmospheric phase
stability.  Another pointing scan is taken at the end of the map so that
the data can be corrected by interpolation in post-processing. The typical
drift is about 10\arcsec\ but can be as large as 1\arcmin\ and is mostly
due to the inaccuracy of the TLE trajectory. In the case of large drift,
we perform an additional pointing scan in the midst of the map so that the
correction can be more accurately interpolated.

Initially, the map scans were performed by moving the telescope in the
elevation direction in attempt to minimize feedarm vibration. However, no
substantial difference was seen between maps taken with elevation and
azimuth scans in tests during a single night in February 2009. As a
result, nearly all of the maps since March 2009 were performed by scanning
along the azimuth direction so that the elevation-dependent Zernike
corrections could be applied by the M\&C system prior to each scan.  Since
these corrections change only when the elevation changes, it is natural
for them to remain constant during each azimuth scan.  In this way, the
holography data are acquired using the same primary surface conditions
with which high-frequency observational astronomical data are obtained.
The servo following error (indicated position minus commanded position) on
these high rate scans ($>\!2$\arcdeg\ minute$^{-1}$) is impressively small
($\lesssim 1$\arcsec) during most of the scan, with a brief oscillation at
the beginning of the rows following a reference scan.

\section{Data Processing and Analysis}

\subsection{Pre-processing}

The holography backend data and the antenna encoder position data are
sampled asynchronously, each at a different rate, and are written to
separate FITS files.  Matlab macros are used to read these files and
produce data plots and diagnostic statistics useful to the observers
during the observing sessions, including phase stability, total power, and
dynamic range on the reference scans.  When the map is completed, another
macro concatenates the data from the hundreds of scans into a set of three
ASCII files.  The first file contains the timestamped azimuth and
elevation position offsets of the antenna with respect to the
instantaneous satellite position.  The second file contains the absolute
azimuth and elevation for the satellite position at the time of the center
of each map row.  The third file contains the timestamped phase and
amplitude computed from the correlator products.  These files are used in
the subsequent regridding and Fourier transform steps. If the measured
pointing corrections are supplied to the Matlab macro, it will compute and
apply interpolated pointing corrections to the data prior to creating the
files.  There is also an option to correct for phase drifts during the
map, but we found this feature to be unnecessary.

\subsection{Regridding and Fourier transform}

The holography post-processing, and much of the subsequent data analysis
and display, are done within the computational framework of the {\it
Mathematica} package \citep{Wolfram10}.  The first step is the calculation
of the direction cosine ($u$-$v$) coordinates associated with each data
sample.  This calculation is based on the instantaneous satellite Az and
El, computed from the orbital elements given in the satellite ephemeris;
the commanded telescope delta-Az and delta-El from the satellite Az and
El; and the telescope encoder Az and encoder El. The relevant expressions
are given by Equations~21--24 of \citet{Rahmat-Samii85}.  Also, a $u$-$v$
plane geometric phase correction, described in \citet{Schwab08}, is
applied to the raw data in order to remove a phase gradient whose presence
is due to the rapidly changing differential geometric delay between the
two holography feeds.  (If the assumed survey position of the reference
feed is in error, a nonlinear residual phase error will result.  A survey
accuracy in reference horn position better than ${\sim}25$~cm is required,
according to the analysis in \citet{Schwab08}.)

In our analysis software no assumption is made that the input data lie on
a regular, rectangular grid.  That is, the sampling distribution could be
arbitrary: a daisy-petal pattern, a raster map, or completely random.  We
use a convolutional gridding scheme \citep[e.g.,][]{Schwab84} to grid the
data prior to application of the fast Fourier transform (FFT) algorithm,
just as in the packages---AIPS, CASA, GILDAS, MIRIAD, etc.---that are
widely used for radio interferometric aperture synthesis data reduction.
The gridding convolution kernel is a two-dimensional tensor product of
(identical) one-dimensional zero-order spheroidal functions, supported on
a square region of width \hbox{$m=8$} grid cells on each side; the same
choice of kernel (except $m=6$) is used as the default choice in AIPS and
CASA\@.

To suppress ringing artifacts, we apply a smooth, circularly symmetric
data taper---i.e., apodization function---with an edge amplitude of about
0.2 (relative to the central amplitude).  We use a prolate spheroidal wave
function taper \citep{PW93}, but a simple Gaussian would be about equally
effective.

We use a $1024\times1024$ input grid with cell sizes $\Delta u=\Delta
v=0.0075$ degrees.  The output, aperture plane grid then has spacings
$\Delta x=\Delta y=\lambda/(1024 \Delta u)\approx0.191$~meters, when we
observe the Galaxy~28 satellite beacon at 11.702~GHz.  The actual spatial
resolution of the maps is ${\sim}0.34$~m---or somewhat larger, depending
on the the choice of the apodization function.

\subsection{Phase unwrapping}

The GBT actuators have a throw of about $\pm2.5$~cm.  At an observing
frequency of 11.7~GHz ($\lambda\approx2.56$~cm) a peak-to-peak surface
error in excess of $\lambda/2\approx1.28$~cm will lead to a phase wrap in
the holography surface error map.  Actual peak-to-peak errors, in the
presence of malfunctioning actuators, can be as large as 5~cm.  Thus as
many as four phase wraps, peak-to-peak, might exist across an 11.7-GHz GBT
holography map.  At the start of our 2009 holography campaign, we
identified twenty to twenty-five inoperable or malfunctioning actuators
(among the 2209 on the GBT)\@.  In our initial, January~4, 2009, map there
were phase wraps in the neighborhood of several of these (six or eight).

Thus, we routinely apply phase unwrapping algorithms in our holography
data processing.  Two-dimensional phase unwrapping is nontrivial,
but---fortunately---the remote sensing and medical-imaging communities
have developed sophisticated and reliable two-dimensional phase unwrapping
algorithms.  Just prior to the launch of the new generation of synthetic
aperture radar satellites used for digital elevation mapping, a book with
state-of-the-art algorithms was published \citep{Ghiglia98}.  This volume
includes C-code implementation of eight different algorithms.  We apply,
routinely, either the Goldstein or the Flynn algorithm, included therein.

\subsection{Defocus fit and fitting for Zernike terms}

B.~Nikolic (private communication) developed an axial defocus model
appropriate to the offset Gregorian geometry of the GBT\@. (Whereas for a
symmetric telescope, this would include only a quadratic term,
higher-order terms are included in the Nikolic model.)  We do a
one-parameter fit to derive the amount of axial defocus.  This amount of
defocus (together with best-fit constant and tilt terms) is removed from
the surface error map prior to further analysis.  We model the remainder
of the large-scale surface error by fitting a Zernike orthogonal
polynomial expansion, typically including the first thirty-six or
fifty-five Zernike terms.

\subsection{Removal of diffraction rings}

\begin{figure}
\plotone{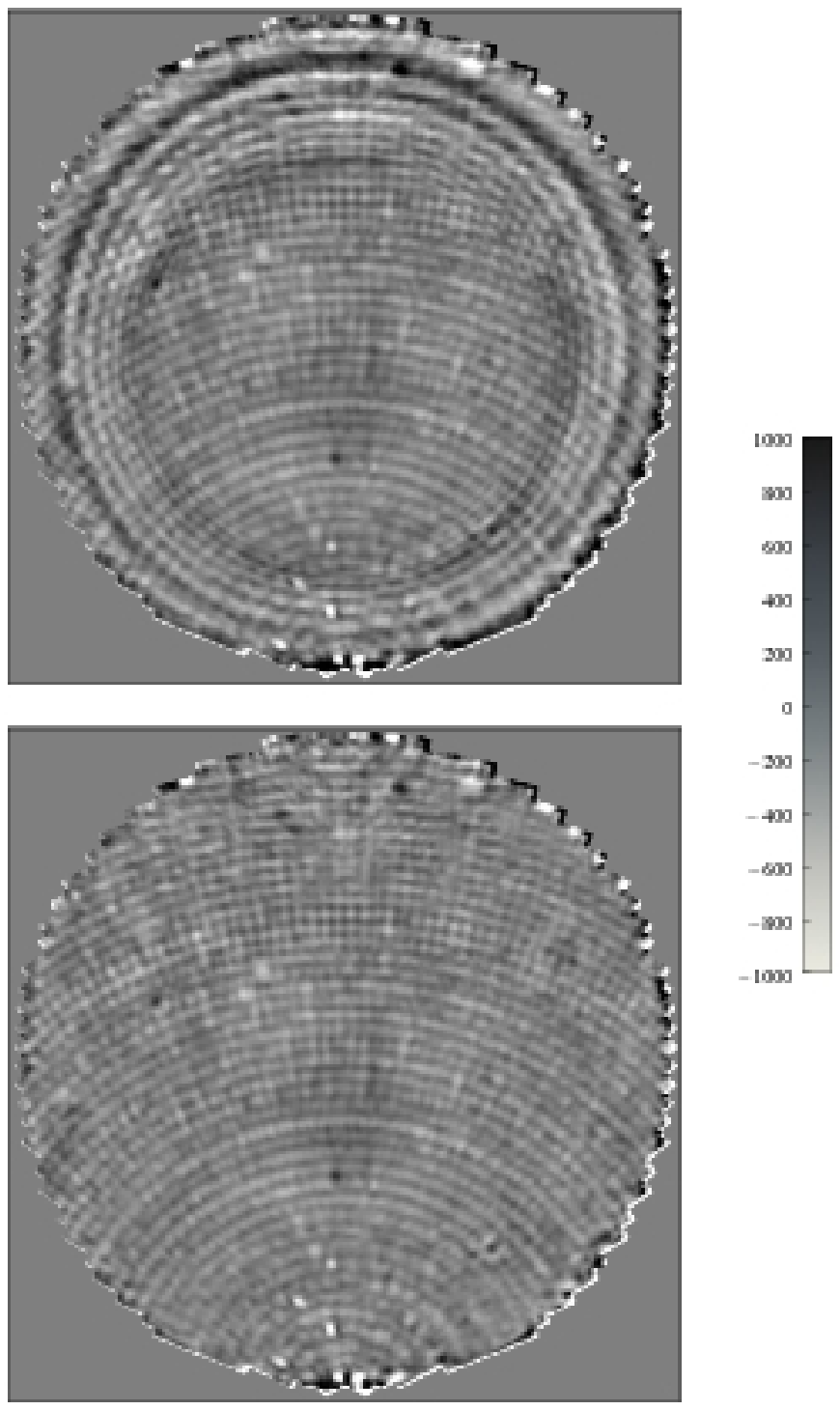}
\caption{\label{diffig1} 
The Sept.~11, 2009, GBT holography surface error map, before {\it(top)}\/
and after application {\it(bottom)}\/ of our diffraction ring filtering
algorithm.  Large-scale structure---as represented by a fifty-five term
Zernike polynomial fit---was removed from the surface error map prior to
diffraction ring removal.  The plot legend is labeled in microns.}
\end{figure}

The main GBT holography receiver is mounted at the Gregorian focus, rather
than at prime focus.  This choice was made primarily because the
high-frequency receivers---i.e., those that require the best surface
figure---also reside at Gregorian focus. Holography maps made from
Gregorian focus are sensitive to the combined effects of surface errors on
both the subreflector and main reflector.  However, a significant drawback
of this configuration for holography is the effect of diffraction that
occurs at the eight-meter diameter subreflector.  In particular, the outer
radii of the holography map phase (and amplitude) distributions are quite
seriously contaminated by diffraction rings due to the subreflector edge;
an example of this, occurring within the phase distribution map from the
September, 2009 holography session, is shown in the upper panel of
Figure~\ref{diffig1}.  We have modeled the diffraction pattern
corresponding to the dual-offset GBT design using the Zemax package for
optical system analysis \citep{Zemax10}.  The simulated amplitude rings
are concentric and circularly symmetric, and the number and spacing of
rings match the observations.  The simulated phase distribution shows the
same pattern, except that it varies circumferentially, and it is in
reasonable accord with our observations.  The peak-to-peak diffraction
ring phase excursion in the Zemax model is approximately $\pm0.4$~radian
at 11.7~GHz, which corresponds to a peak-to-peak surface error
perturbation of about $\pm400\,\mu$m.  This is about the level which is
seen in the holography maps.  Clearly, we must correct for the effect.
Close inspection of our holography images reveals that the diffraction
rings in the aperture plane are non-concentric; in particular, the rings
near the bottom edges of the maps are more closely spaced than those near
the top edges.  Since we also find that the rings shift a bit from one
session to the next, we have adopted an empirical approach to removing
their effects.  We ``filter out'' the rings by: (1)~first removing the
large-scale error, as given by the Zernike fit; (2)~manually tabulating
the positions of points along the peaks of the observed diffraction rings
(here we generally use the amplitude map rather than the phase);
(3)~fitting for quadratic polynomials modeling the $x$-$y$ positions of
the diffraction ring centers, as a function of radial distance from the
center of the aperture plane; (4)~then at each given location in the
holography map, finding the median residual surface error along the
best-fit circumferential diffraction ring arc (of angular size, say,
$\pi/4$ or $\pi/8$ radians) centered on that point; (5)~subtracting from
the map the errors calculated in Step~4; and finally, (6)~optionally
adding the Zernike fit back into the surface error map.

\begin{figure}
\plotone{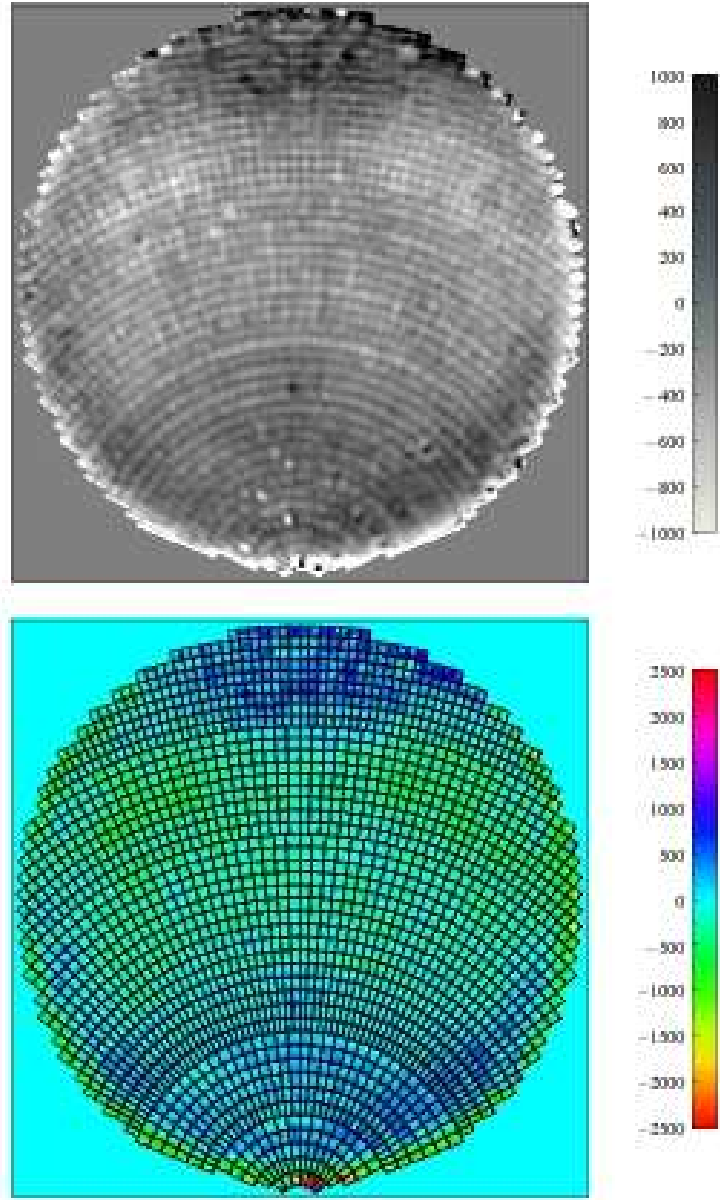}
\caption{\label{diffig2} 
{\it(Top)}\/ The Sept.~11, 2009, GBT holography surface error map,
ring-filtered and Zernike-restored (i.e., the same as Figure~2b, but
with the fifty-five term Zernike polynomial model added back in).
{\it(Bottom)}\/ Same as above but displayed in color, with the GBT panel
layout superimposed.  The plot legends are labeled in microns.}
\end{figure}

Figure~\ref{diffig1} shows our Sept.~11, 2009, Zernike-removed surface
error map before and after diffraction ring removal.  Figure~\ref{diffig2}
shows that map with the Zernike model restored.

\section{Results}

\subsection{Surface maps and adjustments}

In our analysis we have no way of separating subreflector surface errors
from main reflector errors.  The manufacturing errors of the subreflector
panels were held to under $<50\,\mu$m r.m.s., and the achievement of the
$100\,\mu$m r.m.s.\ overall surface accuracy specification for the
subreflector---including panel setting---was verified by photogrammetry
before telescope commissioning ({\it ca.}\ 2000).  The contemporaneous
photogrammetric survey measurements of the primary surface actuator
control points (in their home positions) had, on the other hand, r.m.s.\
errors of order twice as large according to estimates provided by the
contractor \citep{GSI2011}.  Our working assumption is that primary
surface errors are dominant.  In principle, though, the contribution due
to large-scale surface errors on the subreflector can be corrected by
adjusting the main reflector.

In the typical holography map, surface features on the primary mirror as
small as 0.5~m are visible, compared to the typical panel size of 2~m by
2.5~m.  A number of large features coincident with specific actuators were
immediately identified, and soon traced to electrical problems either with
the actuator motors, their position sensors, or the associated cabling.
These problems were repaired during the following months as the campaign
continued through several iterations of holography mapping, surface
adjustments, and radiometric testing at 90~GHz with the MUSTANG bolometer
camera \citep{Dicker08}.

The surface adjustments take the form of adjustments to the actuator home
positions stored in a telescope configuration file.  We have also
performed manual adjustments of the panel mounting screws at two sets of
panel corners which showed particularly noticeable errors not correctable
by motion of the associated actuator, each of which are shared by (up to)
four panels.

\begin{figure}
\plottwo{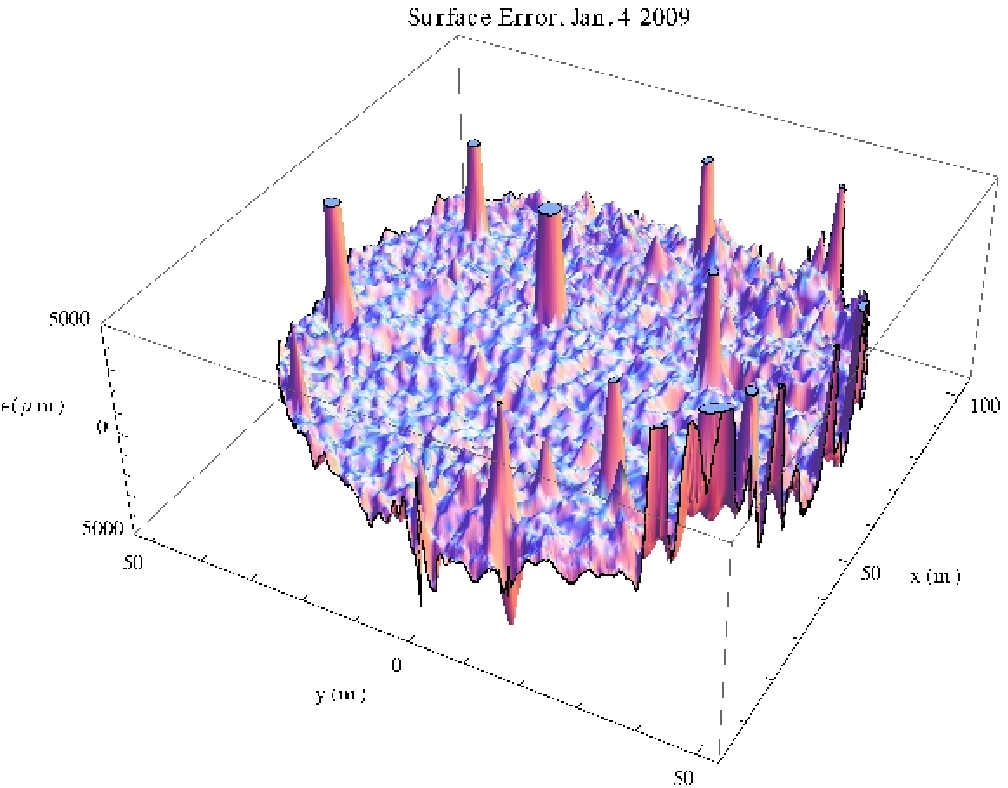}{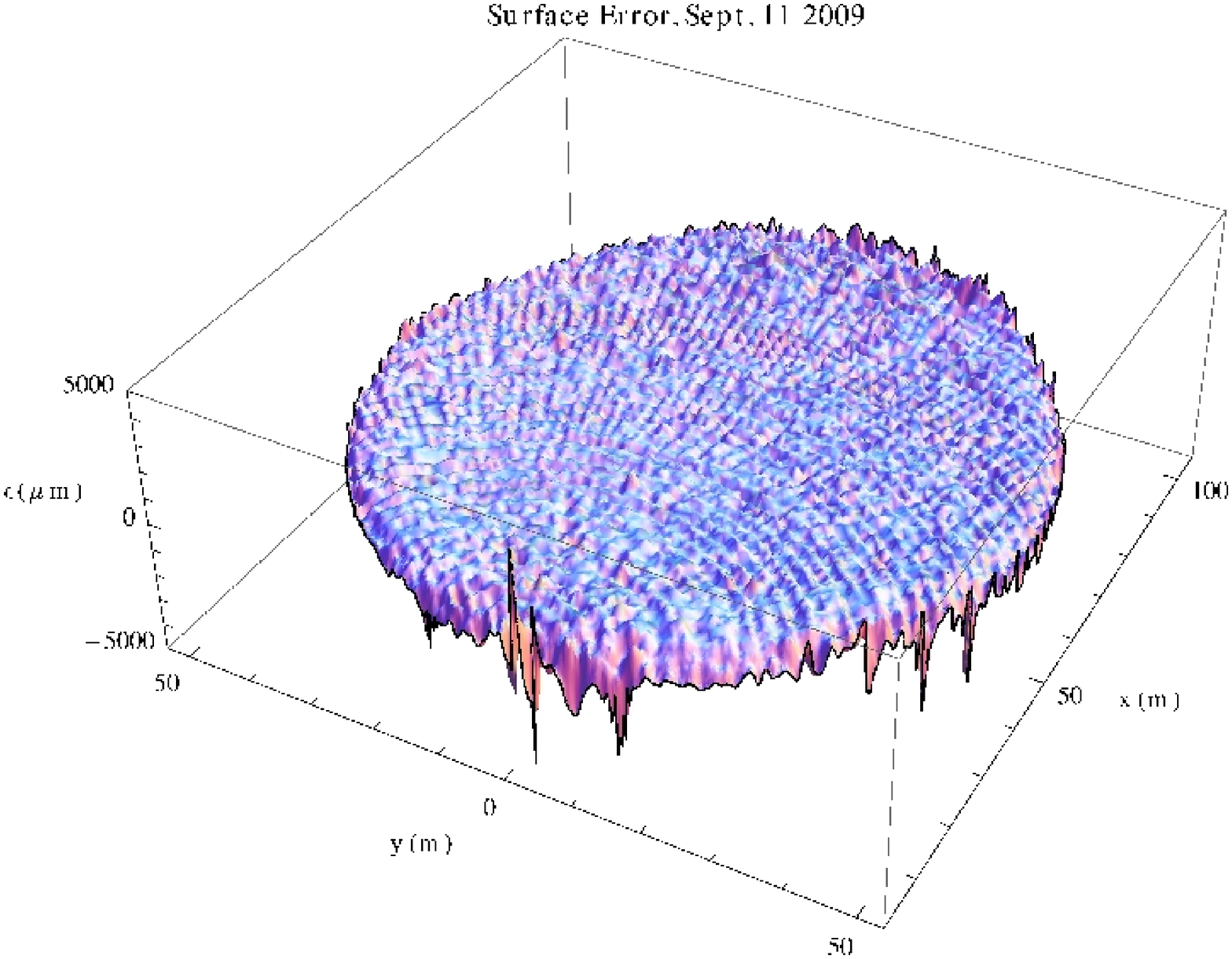} 
\caption{\label{JanSeptComparison0} 
Surface error map derived from the January 4, 2009 holography data
{\it(left)}, compared with the September 11, 2009 map {\it(right)}, each
shown in three-dimensional perspective.  Most of the gross outliers which
are evident in the January map occur in the regions of influence of
malfunctioning actuators.  However, a few of the outliers on the outer
edge of the dish are due to shimming adjustments made to actuator mounting
points following the photogrammetric survey performed in June of 2000. An
actuator which is far out of position will result in image artifacts
(e.g., ripple effects) that contaminate the derived surface map well
beyond the region of mechanical influence of the problem actuator.  All
2209 primary surface actuators are believed to have been functioning
correctly during the September 11, 2009 observation session.}
\end{figure}

Our first successful high-resolution GBT holography map was acquired on
January 4, 2009.  The surface error map derived from this session is shown
in the left-hand panel of Figure~\ref{JanSeptComparison0}.  A large number
of outliers are evident in this three-dimensional perspective display.
Most of the gross outliers occur in the regions of influence of
malfunctioning actuators.  A few of the outliers on the right-hand, outer
edge of the dish are due to shimming adjustments which were made following
the photogrammetric survey performed at the end of the telescope
construction period, in June 2000. Grossly malfunctioning actuators can
result in image artifacts (e.g., ripple effects) which contaminate the
derived surface map well beyond the area of mechanical influence of any
particular problem actuator.

Most of the malfunctioning actuators were repaired during the late winter
and the spring months of 2009.  Some actuator problems were intermittent
and took relatively longer to diagnose; however, all were fully functional
by the end of summer.  Physical access to actuators, which is via
cable-supported catwalks, is time-consuming and sometimes logistically
difficult.  However, the problems due to shimming cleared up immediately,
following the application of the first high-resolution holography-based
actuator zero-point corrections based on the January~4 holography map, as
was verified by the first follow-up holography session on February~9.

\subsection{Surface accuracy improvements}

Twenty-one successful holography mapping sessions were conducted between
February and September, 2009.\footnote{See {\tt
https://safe.nrao.edu/wiki/bin/view/GB/PTCS/TraditionalHolographyProject}
for complete details, including tabular summary statistics.}  The first
couple of iterations of mapping and surface adjustments produced marked
improvements in surface accuracy.  Following that, further substantial
improvements followed only from holography data acquired under the most
favorable meteorological conditions.  Each high-resolution map required
three to four hours of continuous observation.  The most useful maps (for
surface setting) were acquired only during the dead of night, during
conditions of stable ambient temperature, near thermal equilibrium (i.e.,
after the daytime thermal effects had damped out; ideally inside a period
of a few days prevailing overcast), low winds, and good telescope
pointing.  Data obtained during sub-optimal conditions (e.g.,
post-sunrise, as well as clear, cold night) were useful also, particularly
in recognizing the repetitive patterns of thermal-gradient induced and
gravitationally induced systematic panel distortions (see \S5.4).

In the presence of extreme outliers due to malfunctioning actuators, we
found the raw r.m.s.\ statistic not to be generally useful as a summary
measure of surface accuracy, because in this case it is not reflective of
``typical'' surface roughness and is not a reliable predictor of
radiometric-equivalent surface accuracy.  However, we found the median
absolute deviation about the median residual, commonly known as the MAD
estimator \citep{Hampel86}, to be highly useful.  For a normal distribution,
the MAD is 1.4826 times the standard deviation.  In our work, we found
found the raw median absolute deviation surface statistic, scaled by
1.4826, to be in relatively close agreement with a ``masked'' root-mean
square statistic, where ``masking'' implies giving zero weight to all grid
points within, say, a 2.5-m radius of any problem actuator.

In order to predict the radiometric efficiency for any standard GBT
Gregorian receiver it is also useful to compute weighted r.m.s.\
statistics,\footnote{Or rather, the masked r.m.s.\ (or scaled MAD) if
there are serious outliers.} with the weights corresponding to the
standard feed illumination taper of $-13$~dB\@.  For comparison with
MUSTANG measurements, it is useful to compute an unweighted r.m.s.\ over
the central 90-m diameter of the dish, corresponding to the Lyot stop of
the MUSTANG optics. (MUSTANG does not use a tapered illumination pattern.)

Because of the effects of the diffraction rings, which were discussed in
\S4.5, we made slower progress in improving the outer portion of the
surface than the inner portion.  As we proceeded we found the combination
of an inner MAD (or masked r.m.s.) statistic (computed, say, for radii
$r<28$~m) and the corresponding ``outer'' statistic (for $r>28$~m) to be
useful in measuring the relative progress in improving both parts of the
dish.  (We chose $r=28$~m because that was the smallest radius into which
we were generally able to distinguish individual rings for the empirical
diffraction ring fitting; we also tabulated the statistics for $r=36.7$~m,
because half of the actuators are within that radius and half are
outside.)

\begin{figure} 
\includegraphics[height=434pt]{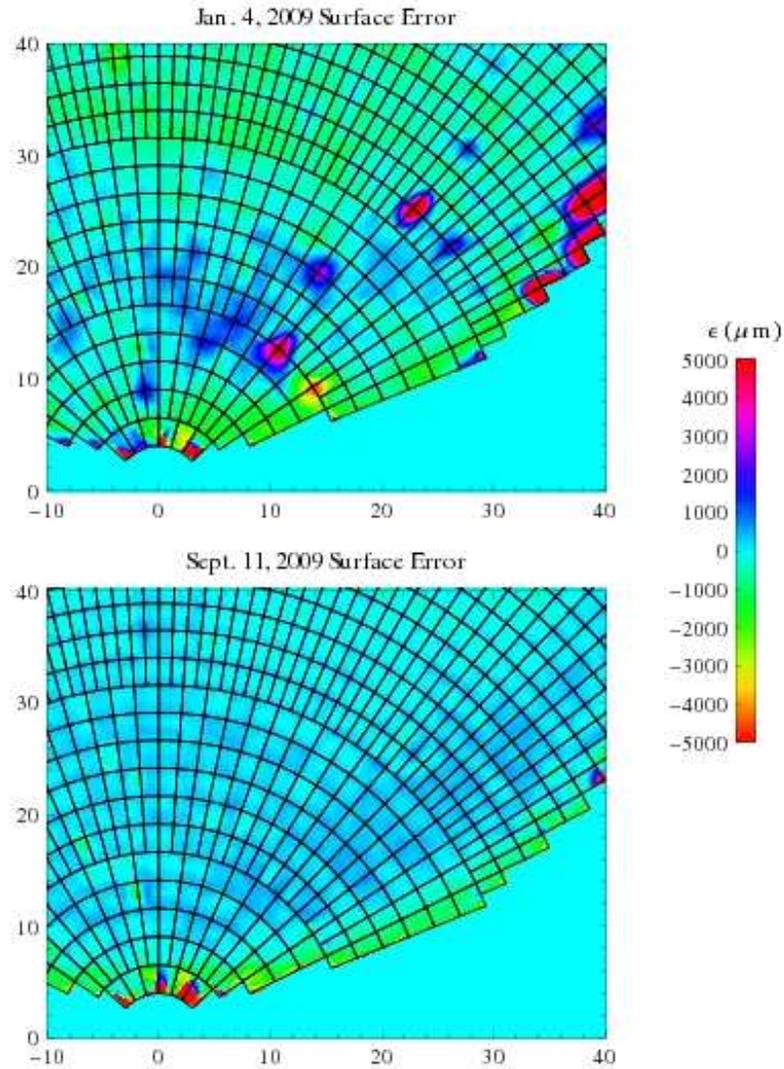}
\caption{\label{JanSeptComparison1}
Detailed comparison of the surface error maps derived from January 4 and
September 11, 2009 holography data, showing a portion of the surface near
the paraboloid vertex.  At top, the brightest two red interior globular
``blobs'' (in Hoops~6 and~13, counting from the vertex) are prototypical
examples of the actuator influence functions at transition tiers (Hoop~6,
the blob appearing pear-shaped) and at regular panel intersections
(Hoop~13, the blob radially elongated).  The persistent features in Tier~1
(similar in both maps) are due to permanent damage from ice that has
fallen from the feed arm, and to incidents which involved mechanical
interference with the access platform.  Additional ice damage is present
in Tiers~2, 4, and~6 and is evident in the holography maps when a more
sensitive transfer function is used for the display.}
\end{figure}

\begin{figure}
\plottwo{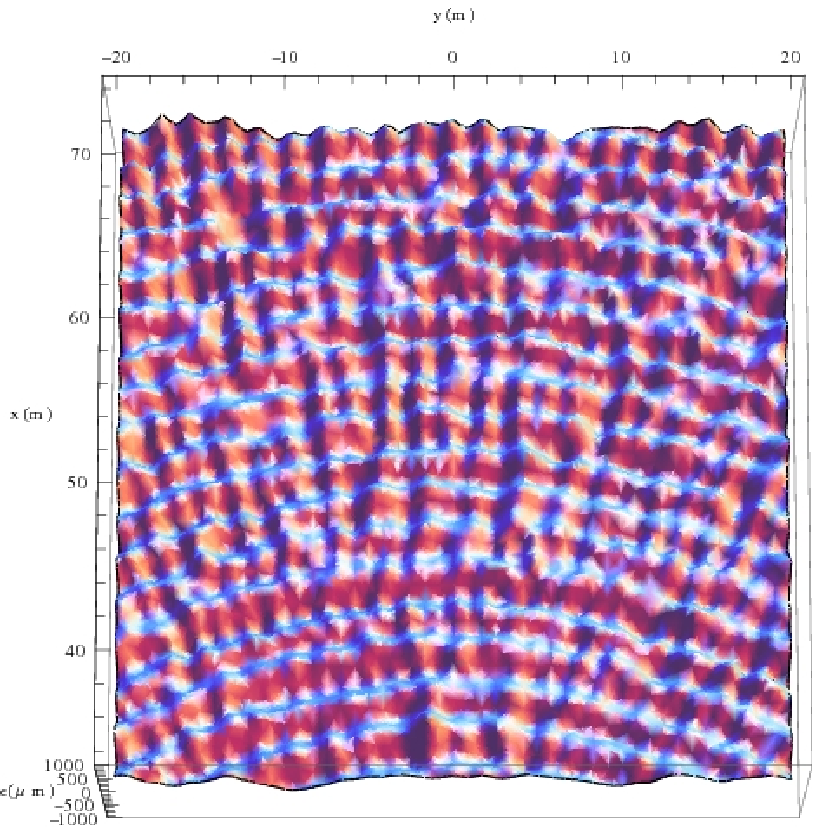}{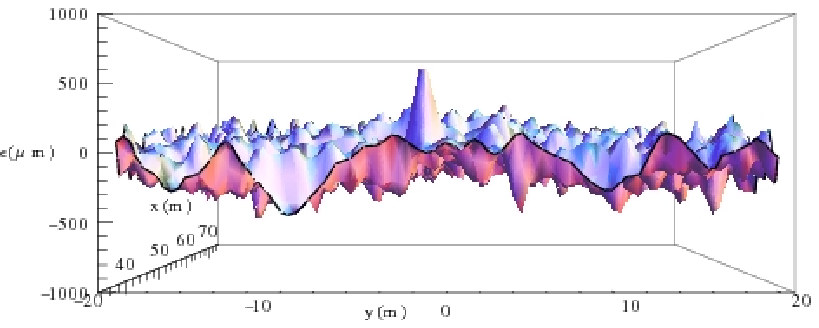}
\caption{\label{Sept11MapCenter}
Three-dimensional perspective plots showing---in greater detail than
Figures~3 and~4---the central $40{\rm\,m}\times40{\rm\,m}$ region of the
September 11, 2009 surface error map.  This area includes the central
portions of seventeen consecutive panel tiers (Tiers 13 through 29).
These tiers can easily be distinguished via the trough-like patterns of
thermal and gravitational sag, which are discussed in \S5.4.}
\end{figure}

The right-hand panel in Figure~\ref{JanSeptComparison0} shows the surface
error map from September 11, 2009, near the end of our holography
campaign.  A more detailed, localized comparison of the beginning and end
results is shown in Figure~\ref{JanSeptComparison1}, in which an overlay
is superimposed to show the panel layout.  Figure~\ref{Sept11MapCenter}
shows detailed perspective plots of the central portion of the September
11, 2009 map.

\begin{figure}
\plotone{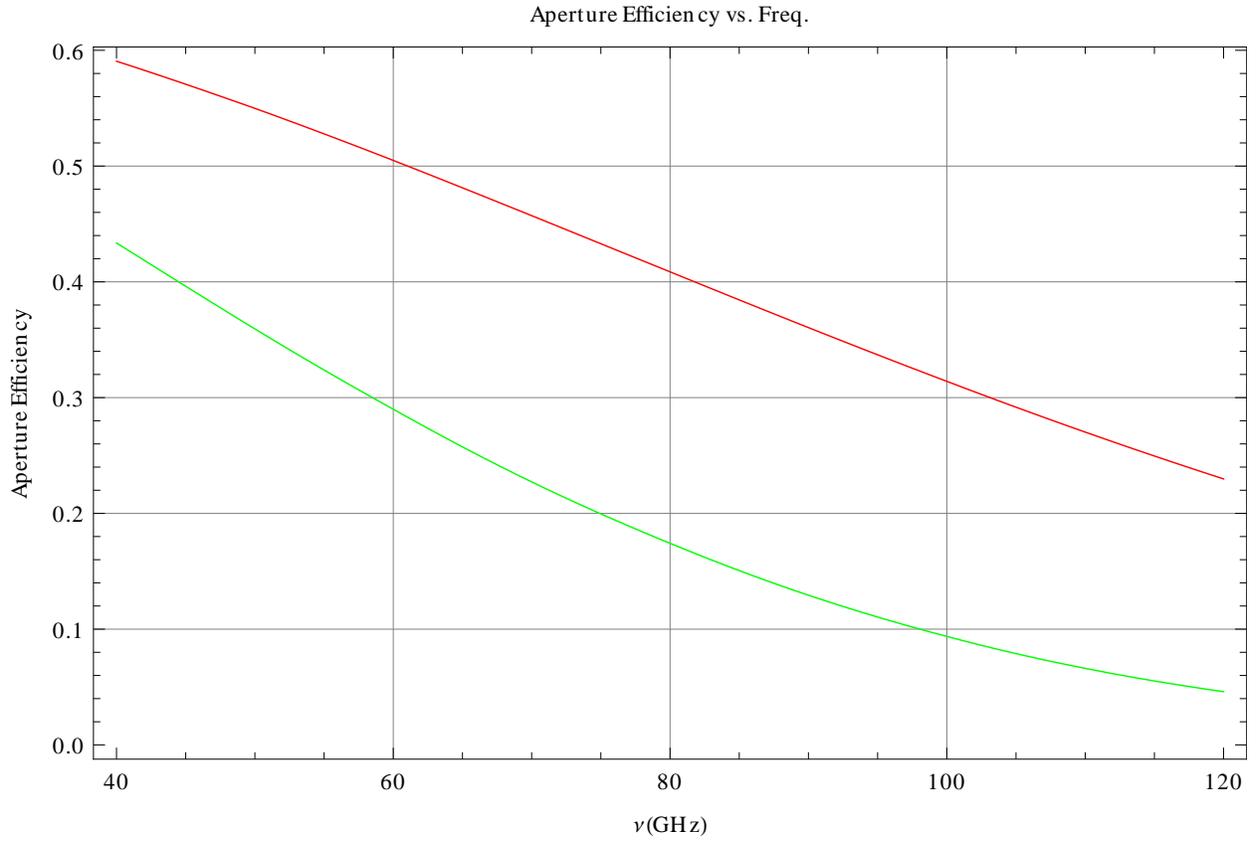}
\caption{\label{ApEffPlot} 
Predicted aperture efficiency vs.\ frequency curves for the GBT with
the surface as measured on January~4, 2009 (lower curve, green), and
September~11, 2009 (upper curve, red), 
assuming illumination of the full 100-m primary
aperture by a feed with the standard $-13$~dB edge taper.}
\end{figure}

Figure~\ref{ApEffPlot} shows the aperture efficiency vs.\ frequency curves
that we would predict based on the January~4 and September~11, 2009 GBT
holographically measured surfaces, for astronomical observations using a
receiver with the standard $-13$~dB edge taper. These predictions are
compared with actual radiometric observations in the next section.  To
calculate the gain, we fit an interpolation function to the measured
(phase-unwrapped) holography surface error distribution and performed the
aperture-plane integration defined by Equation~1 of \cite{Ruze66}. The
peaks in the gain curves occur at 75.2~GHz (91.9~dB) and 119.4~GHz
(96.6~dB) for the January~4 and September~11 surfaces, respectively.  For
the aperture-efficiency calculation we included a diffraction loss based
on geometric theory of diffraction (GTD) calculations by
\cite{Srikanth94}.

A complete list of successful nighttime holography sessions, representing
the data acquired under the best observing conditions, is given in
Table~\ref{rmstab1}.  The final column in the table shows the r.m.s.\
surface error (in the normal direction) estimated from the holography map
(diffraction ring corrected, and low-order Zernike polynomial terms
removed), with the r.m.s.\ computed over the entire aperture ($r<50$\,m),
excluding points within a 2.5-m radius of any grossly malfunctioning
actuators.  The number of such actuators is shown in Column~3.  By
midsummer 2009, all the grossly malfunctioning actuators had been
repaired.  Steady progress was made in improving the surface between
January and May of 2009.  In the January 2010 map, the observed r.m.s.\
error is in the same range as that of the summer maps.  The mean r.m.s.\
for the last four sessions is 223\,$\mu$m.  Table~\ref{rmstab2} shows the
results from two daytime holography mapping sessions.  For the daytime
sessions the mean r.m.s.\ is 319\,$\mu$m, which is 44\% greater than that
of the final four nighttime sessions.

The nighttime r.m.s.\ value implies a total surface r.m.s.\ of
${\sim}240\,\mu$m, assuming that the residual large-scale component of
surface error is ${<}100\,\mu$m r.m.s.  This assumption should generally
be valid during stable thermal conditions but can be more routinely
guaranteed by measuring and compensating for large-scale errors by executing
the automatic OOF holography procedure (``AutoOOF'') prior to observations
\citep{Hunter2009}.

\subsection{Efficiency improvements}

Following each round of holography measurements, typically within a few
days or weeks, we tested the new surface corrections radiometrically using
MUSTANG\@.  The active surface software allows one to quickly load a new
configuration file containing the complete list of actuator position
offsets.  Taking full advantage of this capability, we acquired small
images of a bright point source quasar using a ``daisy petal'' scan
pattern, which repeatedly passes through the target position with a
steadily advancing sequence of angles of attack.  We first acquired an
image with the old surface, then with the new surface, and repeating the
sequence a few times.  The purpose of this test was to confirm the
expected increase in the response of the detector as the dish was made
more accurate.  Prior to each test, we performed OOF holography in order
to first measure and correct the surface for the current large-scale
deviations (mostly thermal in nature).  By doing this step, we could
ensure that the ``old'' surface produced its best signal, and thus any
improvement subsequently seen with the ``new'' surface meant that the
smaller-scale surface errors had been reduced.  As soon as the improved
performance of the ``new'' surface was confirmed, it was installed for
general use by astronomers.  As such, several astronomy projects have been
able to take advantage of some or all of these surface improvements,
especially those using MUSTANG\@.  While many more projects in the current
(2010--2011) season are underway, data from some MUSTANG projects from the
2009--2010 season have already been published
\citep{Mason10,Korngut10,Shirley11}.

\begin{figure}
\plotone{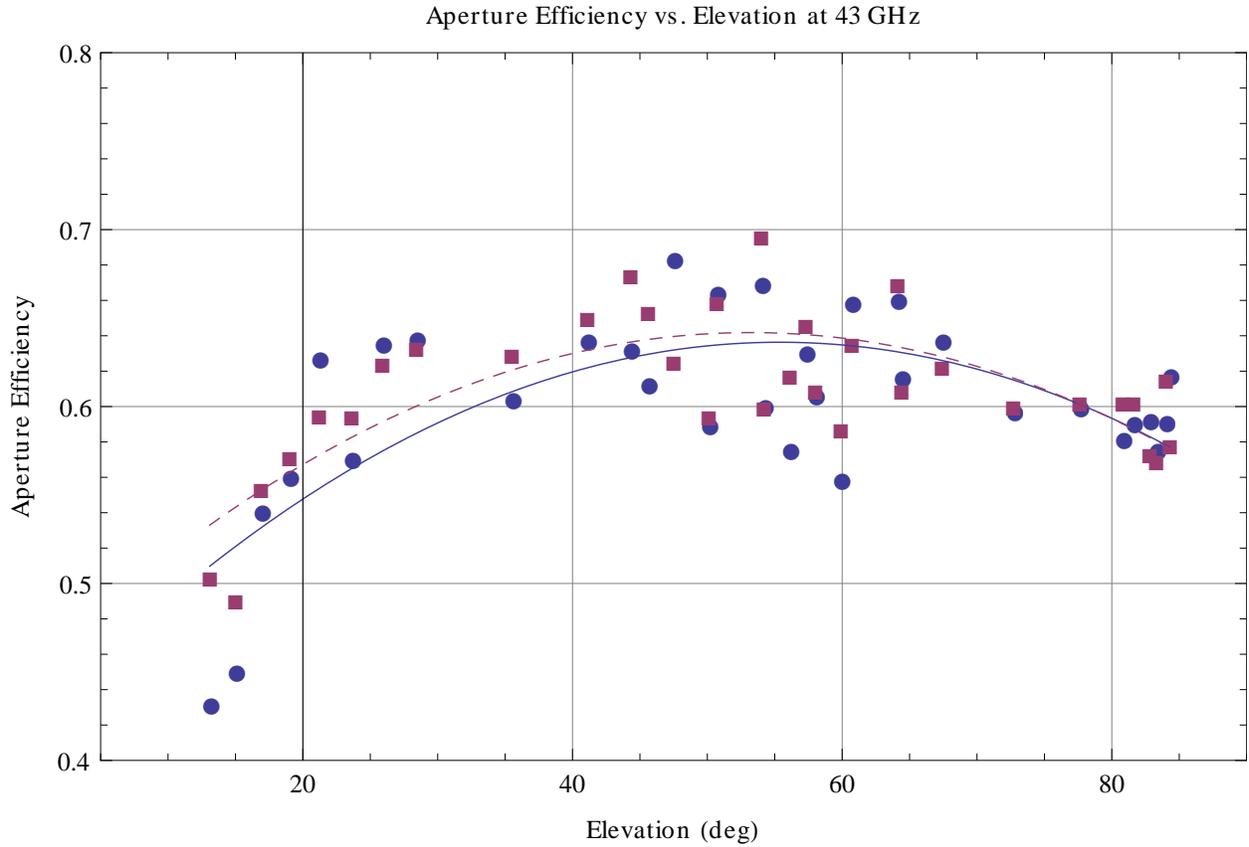}
\caption{\label{MeasuredApEff} Plot of GBT aperture efficiency vs.\
elevation measurements taken at 43.1~GHz on October~4, 2009.
Left-circular polarization data are represented by the filled circles
(shown in blue); right-circular polarization data by the filled squares
(red).  Quadratic fits to the data are shown by the solid (blue) and
dashed (red) curves, for left- and right-circular polarizations,
respectively.}
\end{figure}

During these radiometric tests, we often interspersed measurements of the
same quasar using the original January~4, 2009 surface in order to assess
the total improvement in the surface since the initiation of the
holography campaign.  However, because there were many failed actuators in
the January~4, 2009 surface, the resulting performance ratio increasingly
became an underestimate as these actuators were fixed. We thus decided to
construct a surface file which mimicked all the measured large errors due
to failed actuators.  The total improvement ratio measured in this manner
on November~23, 2009 was 2.4 with MUSTANG (90~GHz) and 1.4 with the Q-band
receiver (43~GHz).  Using the best surface, a gain curve was measured from
15\arcdeg\ to 85\arcdeg\ elevation at 43~GHz on October~4, 2009 using 3C48
and 3C147 (see Figure~\ref{MeasuredApEff}).  A second-order polynomial fit
to the data showed a nearly flat curve with a maximum aperture efficiency
of 63\%, as compared with ${\sim}45\%$ measured several years ago, and
consistent with the 1.4 improvement factor. Also, these values imply an
improvement in the total surface error from $390\,\mu$m \citep{Nikolic07b}
to ${\sim}240\,\mu$m rms, which is consistent with the improvement in the
measured surface errors in the holography maps themselves.  Based on this
result, we predict the aperture efficiency at 90 GHz is between
$30$--35\%.  Aperture efficiency measurements with MUSTANG are complicated
because this term is degenerate with the receiver optical efficiency;
however the best estimates are in agreement. We note that MUSTANG
illuminates the telescope surface differently from the other GBT receivers
because it has a cold Lyot stop corresponding to a diameter of 90\,m,
rather than the standard $-13$~dB Gaussian taper at a diameter of 100\,m.
Thus, although the total illumination efficiency is similar between these
instruments, the effective aperture efficiency may differ depending on the
distribution of surface errors as a function of radius.  In any case, the
aperture efficiency prediction will be tested soon using the new W-band
feedhorn receiver (68--92~GHz) currently under development
\citep{Frayer2010}.

\subsection{Modeling of panel-scale effects}

\begin{figure}
\includegraphics[height=450pt]{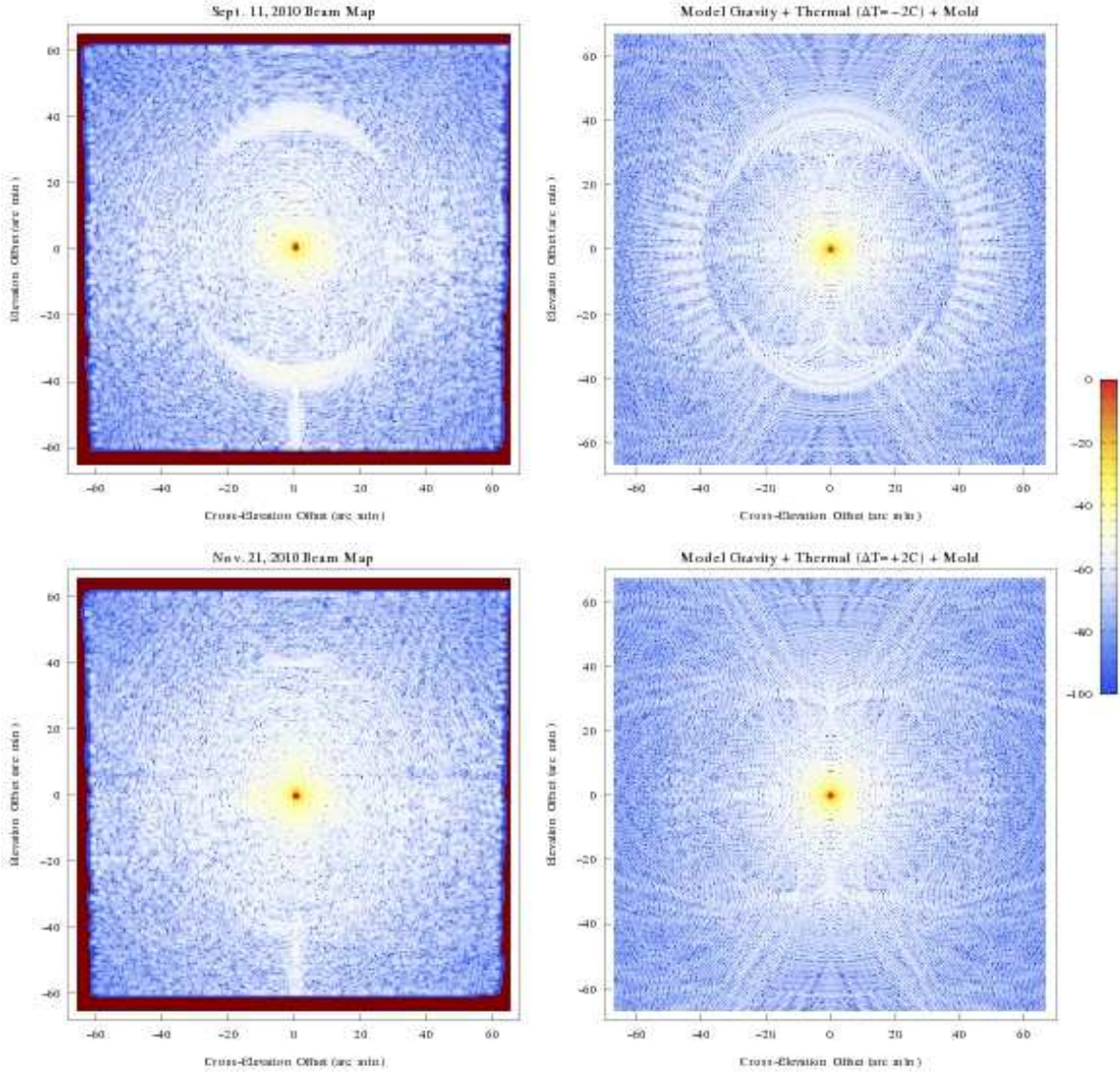}
\caption{\label{ModelComparisonFigure} 
A comparison of observed and model beam patterns: {\it(Upper row)}\/ The
September~11, 2009 map, on the left, is compared with a model which
includes gravitational (elev.\ $44^{\circ}$), thermal gradient ($\Delta
T=-2$\,C), and manufacturing mold errors. {\it(Bottom row)}\/ The
November~21, 2009 daytime map is compared with the model for thermal
gradient, $\Delta T=+2$\,C.  Power in dB (relative to the peak) is
represented by the plot legend on the right.}
\end{figure}

The improvement in the surface accuracy of the primary mirror has had a
correspondingly beneficial effect on the telescope beam pattern.  As a
by-product of each holography map, we obtain a high-dynamic range image
($>60$~dB) of the amplitude pattern.  As the surface was improved, the
level of scattered power was reduced and brought into the main beam.  A
pair of partial arc-like features began to emerge in the pattern, and they
are distinctly evident in the September~11, 2009 map shown in the upper left
panel of Figure~\ref{ModelComparisonFigure}.  Because the features appear
at a radius of about 0.6~degrees, they must be associated with surface
structure on the scale of individual panels ({\it cf.}~\cite{Greve2005},
\cite{Greve2010}, \cite{GrevBrem2010}).  We reviewed the results of a
finite-element analysis of the surface panel design, performed by the
manufacturer, who modeled the panel response to gravity and to a
front-to-back temperature gradient of $2^{\circ}$\,C\@.  Each of these
effects produces a systematic effect on the shape of the panel.  While
gravity produces a sag proportional to the projected gravity vector,
temperature gradients produce a sag or bulge depending on the sign of the
gradient \citep{vonHoerner71}.

\begin{figure}
\includegraphics[height=450pt]{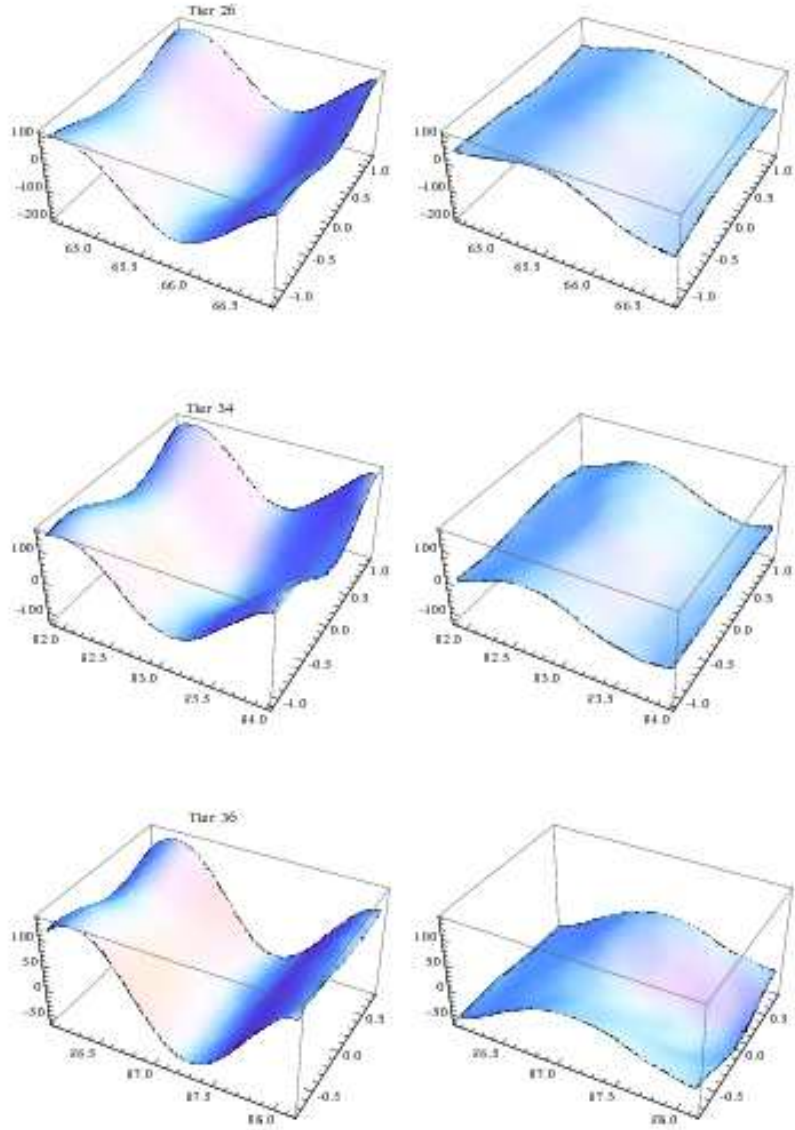}
\caption{\label{TierAvgPanels} 
Comparison of tier-average surface error profiles from the September~11,
2009 map, shown at left, and the November~21, 2009 map, at right.  The
mean gradient $\overline{\Delta T}$ (measured at Tier~16 panel node
16+000) was $-2.69\,{}^{\circ}$C for the September~11 observations and
$+5.00\,{}^{\circ}$C on November~21.  The vertical axis is labeled in
microns; units along the $x$- and $y$-axes are in meters. The selected
tiers show the general behavior: negative $\Delta T$ producing a panel
sag; and positive, a flattening or a bulge. Results for all tiers are
given by \cite{SchwHunt10}, together with results from two other
holography sessions.}
\end{figure}

We were able to scale the models to the geometry of each panel of the
dish.  Next, we assembled the model panels into a complete model of the
aperture, including systematic panel distortions \citep[for full details,
see][]{SchwHunt10}.  Fourier transformation of this model yields the
predicted far-field beam pattern, as shown in the upper right panel of
Figure~\ref{ModelComparisonFigure}.  Both the upper and lower arcs at
0.6~degrees and the fainter arcs to the left and right at larger radii are
qualitatively reproduced by the model.  The September~11, 2009 map was
obtained under clear nighttime skies, when the measured panel gradient
(from surface to rear ribs) was negative ($-2.69^{\circ}$\,C) due to
preferential radiative cooling toward the sky.  The gradient was estimated
using two structural thermometers---one attached to a panel skin in
Tier~16 and the other on the backup structure adjacent to this panel. The
map of November~21, 2009 (lower left panel of
Figure~\ref{ModelComparisonFigure}) was acquired during morning hours from
8:30\,AM--12:30\,PM, local time, when the measured gradient was positive
($+5.0^{\circ}$\,C) because the the majority of panels were being heated
by sunlight. The arcs in the pattern are suppressed. We extracted
tier-averaged profiles from the holography surface maps which clearly
reveal the change from sagging panels to flat or slightly bulging panels
(see Figure~\ref{TierAvgPanels}).  We conclude that a positive temperature
gradient leads to a mechanical deformation in the panels which serves to
counteract the sag due to gravity.  However, daytime heating effects lead
to larger pointing errors and induce large-scale deviations in the dish
which must be actively measured and compensated with OOF holography
\citep{Hunter2009}.  Nevertheless, it is clear that the best nighttime
performance of the dish will be obtained when the panel temperature
gradient is minimal, which generally occurs during periods of overcast
conditions.  Unfortunately, these conditions tend not to be the best for
millimeter-wavelength transparency.

\section{Summary}

We have successfully designed and installed a 12-GHz holography system on 
the GBT\@.  We have performed a campaign of high-resolution holographic 
imaging of the telescope aperture that has enabled us to achieve a 
dramatic reduction in the surface error of the primary mirror.  The 
expected improvements in aperture efficiency have been confirmed.  The 
holography images also reveal the magnitude and direction of the response 
of the panels to different thermal conditions. The holography system is 
fully integrated into the GBT control system and continues to be in 
regular use to monitor the health of the individual actuators, which is 
essential to support high-frequency observations.

\acknowledgments

We thank D. Emerson, B. Nikolic, R. Hills, and T.~K. Sridharan for helpful
advice at various stages of the project.  We would like to thank the referee,
Dr.~Jacob Baars, for his careful review and valuable suggestions.  The
National Radio Astronomy Observatory is a facility of the National Science
Foundation operated under cooperative agreement by Associated
Universities, Inc.

{\it Facilities:} \facility{GBT}

\clearpage

\begin{table}
\caption{Successful nighttime holography mapping observations\label{rmstab1}}
\begin{tabular}{lccc} 
\tableline
\tableline
UT Date     & UT Start Time & No.\ Bad Act. & Trimmed r.m.s.\ ($\mu$m)\\
\tableline
2009 Jan 04 & 05:57 & 21 & 423\\
2009 Feb 06 & 05:24 &  6 & 424\\
2009 Mar 15 & 08:46 &  4 & 321\\
2009 May 27 & 00:13 & 12 & 214\\
2009 Aug 16 & 05:18 &  0 & 210\\
2009 Sep 11 & 03:42 &  0 & 226\\
2010 Jan 21 & 08:33 &  3 & 239\\
\tableline
\end{tabular}
\end{table}

\begin{table}
\caption{Daytime holography mapping observations \label{rmstab2}}
\begin{tabular}{lccc} 
\tableline
\tableline
UT Date     & UT Start Time & No.\ Bad Act. & Trimmed r.m.s.\ ($\mu$m)\\
\tableline
2009 Nov 21 & 13:24 &  2 & 309\\
2010 May 27 & 11:22 &  4 & 329\\
\tableline
\end{tabular}
\end{table}


\begin{thebibliography}{}

\bibitem[Baars et al.(2007)]{Baars07} Baars, J.~W.~M., Lucas, 
  R., Mangum, J.~G., \& Lopez-Perez, J.~A. 2007, IEEE Antennas \& 
  Propagation Mag., 47(5), 24; also arXiv:0710.4244 

\bibitem[Balasubramanyam et al.(2009)]{Balasubramanyam09}
  Balasubramanyam, R., Venkatesh, S., \& Raju, S.~B. 2009,
  in ASP Conf. Ser.\ 407, 
  The Low-Frequency Radio Universe,
  ed.\ D.~J. Saikia, D.~A. Green, Y.~Gupta, \& T.~Venturi,
  (San Francisco: ASP), 434

\bibitem[Bennett et al.(1976)]{Bennett76} Bennett, J.~C., Anderson, A.~P.,
  McInnes, P.~A., \& Whitaker, A.~J.~T.\ 1976, IEEE Trans.\ Antennas \&
  Propagation, AP-24, 295

\bibitem[Constantikes(2004)]{Constantikes04} Constantikes, K.\ 2004, 
  in ASP Conf.\ Ser.\ 314,
  Astronomical Data Analysis Software and Systems XIII, 
  ed.\ F.~Ochsenbein, M.~G. Allen, \& D.~Egret, 
  (San Francisco: ASP), 689 

\bibitem[Constantikes(2007)]{Constantikes07} Constantikes, K.\ 2007,
  USNC/CNC/URSI North American Radio Science Meeting in Ottawa, Canada,
  July 2007

\bibitem[Dicker et al.(2008)]{Dicker08} Dicker, S.~R., Korngut, P.~M.,
  Mason, B.~S, Ade, P.~A.~R., Aguirre, J., Ames, T.~J., Benford, D.~J.,
  Chen, T.~C., Chervenak, J.~A., Cotton, W.~D., Devlin, M.~J., 
  Figueroa-Feliciano, E., Irwin, K.~D., Maher, S., Mello, M., 
  Moseley, S.~H., Tally, D.~J., Tucker, C., \& White, S.~D. 2008, 
  Proc.\ SPIE, 7020:702005

\bibitem[Frayer et al.(2010)]{Frayer2010} Frayer, D. T., O'Neil, K., 
  Lockman, J., Hunter, T., Wootten, A., Maddalena, R., Ghigo, F.,
  \& Langston, G., American Astronomical Society Meeting Abstracts, 216,
  \#414.01

\bibitem[Geodetic Systems, Inc.,(2011)]{GSI2011} Geodetic Systems, Inc.,
  2011, Melbourne, FL;  \url{http://www.geodetic.com/about/}
  
\bibitem[Ghiglia \& Pritt(1998)]{Ghiglia98} Ghiglia, D. C. and Pritt,
  M. D. 1998, Two-Dimensional Phase Unwrapping: Theory, Algorithms, and
  Software, (New York: Wiley)

\bibitem[Grahl et al.(1986)]{Grahl86} Grahl, B.~H., Godwin, M.~P., \&
  Schoessow, E.~P.\ 1986, \aap, 167, 390

\bibitem[Greve \& Morris(2005)]{Greve2005} Greve, A. \& Morris, D. 2005,
  IEEE Trans.\ Antennas \& Propagation, AP-53(6), 2123

\bibitem[Greve et al.(2010)]{Greve2010} Greve, A., Morris, D.,
  Pe\~nalver, J., Thum, C., \& Bremer, M. 2010,
  IEEE Trans.\ Antennas \& Propagation, AP-58(3), 959

\bibitem[Greve \& Bremer(2010)]{GrevBrem2010} Greve, A. \& Bremer, M.
  2010, Thermal Design and Thermal Behaviour of Radio Telescopes and
  their Enclosures, (Heidelberg: Springer)

\bibitem[Hampel et al.(1986)]{Hampel86} Hampel, F.~R., Ronchetti, E.~M,
  Rousseeuw, P.~J., \& Stahel, W.~A. 1986, Robust Statistics: The Approach
  Based on Influence Functions, (New York: Wiley)

\bibitem[Hunter et al.(2009)]{Hunter2009} Hunter, T.~R., Mello, M., 
  Nikolic, B., Mason, B., Schwab, F., Ghigo, F., \& Dicker, S. 2009,
  2009 USNC/URSI Annual Meeting, 45

\bibitem[Korngut et al.(2010)]{Korngut10} Korngut, P.~M., Dicker, S.~R.,
   Reese, E.~D., Mason, B.~S., Devlin, M.~J., Mroczkowski, T.,
   Sarazin, C.~L., Sun, M., \& Sievers, J. 2010, arXiv:1010.5494

\bibitem[Lacasse(1998)]{Lacasse98} Lacasse, R.~J.\ 1998, \procspie,
  3351, 310

\bibitem[Maciolek \& Maddalena(2000)]{Maciolek00} Maciolek, A.~A.,
  \& Maddalena, R.~J.\ 2000, \baas, 32, 1555

\bibitem[Mangum et al.(2007)]{Mangum2007} Mangum, J.~G.,
  Emerson, D.~T., \& Greisen, E.~W.\ 2007, \aap, 474, 679

\bibitem[Mason et al.(2010)]{Mason10} Mason, B.~S., Dicker, S.R.,
  Korngut, P.~M., Devlin, M.~J., Cotton, W.~D., Koch, P.~M., 
  Molnar, S.~M., Sievers, J., Aguirre, J.~E., Benford, D., 
  Staguhn, J.~G., Moseley, H., Irwin, K.~D., \& Ade, P. 2010,
  \apj, 716, 739

\bibitem[Nikolic et al.(2007a)]{Nikolic07a} Nikolic, B., Hills, R.~E.,
  \& Richer, J.~S.\ 2007a, \aap, 465, 679

\bibitem[Nikolic et al.(2007b)]{Nikolic07b} Nikolic, B., Prestage,
  R.~M., Balser, D.~S., Chandler, C.~J., \& Hills, R.~E.\ 2007b, \aap,
  465, 685
 
\bibitem[O'Neil et al.(2006)]{Oneil06} O'Neil, K., Shelton, A.~L.,
  Radziwill, N.~M., \& Prestage, R.~M.\ 2006, in ASP Conf.\ Ser.\ 351,
  Astronomical Data Analysis Software and Systems XV, ed.\ C.~Gabriel,
  C.~Arviset, D.~Ponz, \& E.~Solano, (San Francisco: ASP), 719

\bibitem[Parker et al.(2005)]{Parker05} Parker, D.~H., Anderson, R.,
  Egan, D., Fakes, T., Radcliff, B., \& Shelton, J.~W.\ 2005,
  Precision Engineering, 29, 354

\bibitem[Percival \& Walden(1993)]{PW93} Percival, D.~B. and Walden, 
  A.~T. 1993, Spectral Analysis for Physical Applications: Multitaper and
  Conventional Univariate Techniques, (Cambridge: Cambridge Univ.\ Press)

\bibitem[Prestage et al.(2009)]{Prestage09} Prestage, R.~M.,
  Constantikes, K.~T., Hunter, T.~R., King, L.~J., Lacasse, R.~J.,
  Lockman, F.~J., \& Norrod, R.~D.\ 2009, Proc.\ IEEE, 97, 1382

\bibitem[Radford et al.(1996)]{Radford96} Radford, S.~J.~E., Reiland,
  G., \& Shillue, B.\ 1996, \pasp, 108, 441

\bibitem[RSI(1992)]{RSI92} Radiation Systems Inc., 1992, Technical
  Memo 101,\newline
  \url{https://safe.nrao.edu/wiki/pub/GB/PTCS/LoralTechnicalMemos/TM101.pdf}

\bibitem[Rahmat-Samii(1985)]{Rahmat-Samii85} Rahmit-Samii, Y. 1985,
  IEEE Trans.\ Antennas \& Propagation, AP-33, 1194

\bibitem[Ries et al.(2011)]{Ries11} Ries, P., Hunter, T.~R.,
  Constantikes, K.~T., Brandt, J.~J., Ghigo, F.~D., Prestage, R.~M.,
  Ray,~J., \& Schwab, F.~R., 2011, \pasp, to appear;
  arXiv:0710.4244v1

\bibitem[Rochblatt \& Rahmat-Samii(1991)]{Rochblatt91} Rochblatt, D.~J.,
  \& Rahmat-Samii, Y.\ 1991, IEEE Trans.\ Antennas \& Propagation, AP-39,
  933

\bibitem[Ruze(1966)]{Ruze66} Ruze, J. 1966, Proc.\ IEEE, 54, 633 

\bibitem[Schwab(1984)]{Schwab84} Schwab, F. R. 1984, in Indirect Imaging,
  ed.\ J.~A.~Roberts, (Cambridge: Cambridge Univ.\ Press), 333

\bibitem[Schwab(1990)]{Schwab90} Schwab, F. R. 1990, GBT Memo 28,
  NRAO, Green Bank, WV;\newline
  \url{https://safe.nrao.edu/wiki/pub/GB/Knowledge/GBTMemos/GBT\_Memo\_28.pdf} 

\bibitem[Schwab(2008)]{Schwab08} Schwab, F. R. 2008, PTCS Project
  Note~62, NRAO, Green Bank, WV

\bibitem[Schwab \& Hunter(2010)]{SchwHunt10} Schwab, F. R. \&
  Hunter,~T. 2010, GBT Memo.~271, NRAO, Green Bank, WV

\bibitem[Scott \& Ryle(1977)]{Scott77} Scott, P.~F., \& Ryle,
  M.\ 1977, \mnras, 178, 539

\bibitem[Shirley et al.(2011)]{Shirley11} Shirley, Y.~L., Mason,
  B.~S., Mangum, J.~G., Bolin, D.~E., Devlin, M.~J., Dicker, S.~R., \&
  Korngut, P.~M.\ 2011, \aj, 141, 39

\bibitem[Srikanth(1994)]{Srikanth94} Srikanth, S. 1994, 
  GBT Memo.~215, NRAO, Green Bank, WV

\bibitem[von Hoerner(1971)]{vonHoerner71} von Hoerner, S. 1971, 
   Largest Feasible Steerable Telescope Report No.~36, NRAO, Green Bank, WV;
   \url{https://safe.nrao.edu/wiki/pub/GB/PTCS/NLSRTMemos/65M\_36.pdf}


\bibitem[Wolfram Research(2010)]{Wolfram10} Wolfram Research, Inc., 2010, 
Mathematica, Version~8.0, Champaign, IL

\bibitem[Zemax Development Corp.(2010)]{Zemax10} Zemax Development Corp.,
2010, Zemax: Software for Optical System Design, Bellevue, WA;
\url{http://www.zemax.com}


\end{thebibliography}
\end{document}